\documentclass[10pt, twocolumn,aps,tightenlines,superscriptaddress,amsmath,amssymb,amsfonts,noeprint,longbibliography,prb]{revtex4-2}
\usepackage{graphicx}                   % Include figure files
\usepackage{dcolumn}                    % Align table columns on decimal point
\usepackage{bm}                         % bold math
\usepackage{hyperref}                   % add hypertext capabilities
\usepackage[section]{placeins}
\usepackage{epstopdf}
\usepackage[table, dvipsnames]{xcolor}  % use colors, such as {\color{red} I am red}
\usepackage{newtxtext,newtxmath}        % Times New Roman
\newcommand{\ie}{\textit{i}.\textit{e}.}

\usepackage[colorinlistoftodos]{todonotes}

\begin{document}

\title{Re-entrant localization in fractionally charged electron wave packets}
\author{Y. Yin}
\email{yin80@scu.edu.cn}
\thanks{Author to whom correspondence should be addressed}
\affiliation{Department of Physics, Sichuan University, Chengdu, Sichuan, 610065, China}
\date{\today}
\begin{abstract}
    We investigate the localization transition in fractionally charged electron wave packets, which is injected into a quantum conductor by a single voltage pulse with arbitrary flux quantum. We show that the transition is unidirectional for individual electrons or holes. They always undergo a delocalization-to-localization transition as the flux increases. In contrast, the transition of the neutral electron-hole pairs is bidirectional. As the flux increases, the transition can be a localization-to-delocalization transition or vice versa, which is controlled via the long-time tail of the voltage pulse. The localization-to-delocalization transition occurs in the case
    of short-tailed pulses, which decay faster than Lorentzian. In this case, the directions of the transitions for the neutral eh pairs and individual electrons or holes are opposite. Certain localized neutral electron-hole pairs can first evolve into delocalized ones, then split into individual electrons and holes with localized wave functions, which gives a reentrant localization. The delocalization-to-localization transition occurs in the case of long-tailed pulses, which decay slower than Lorentzian. The reentrant localization vanishes in this case, as the directions of the two transitions are the same. It is also absent in the case of Lorentzian pulses, where the localized neutral electron-hole pairs cannot be excited at all.
\end{abstract}
\maketitle

\section{Introduction\label{sec1}}

In the pioneering work in 1967, Anderson predicted the absence of diffusive behavior due to the quenched disorder \cite{Anderson1958}. This implies the existence of the localization-to-delocalization transition in the zero-temperature limit, where the destructive interference of wave functions inhibits electron transport. It was further shown that the wave function does not undergo a unidirectional transition from the delocalized state to the localized state as the disorder increases. A localization-to-delocalization transition can occur above the band edge, where the disorder enhances the quantum tunneling between localized states and hence produces delocalized behaviors \cite{Economou1984, Grussbach1995, deQueiroz2001}. This can lead to a reentrant localization, which can be controlled via the disorder distribution \cite{Bulka1985, Bulka1987}. The reentrant transition has also been found in quasiperiodic chains, which is induced by the competition between the dimerization and quasiperiodic potential strength \cite{Roy2021}.

In a previous work \cite{Yin2025a}, we have shown that the localization-to-delocalization transition can occur in on-demand electron emitters \cite{ivanov_1995_coher, levitov_1996_elect, dubois_2013_integ, hofer_2014_mach, glattli_2017_pseud}, which is induced by a different mechanism. In this system, individual electrons or holes can be emitted into a quantum conductor driven by a single voltage pulse, which are usually accompanied by a neutral cloud of electron-hole (eh) pairs \cite{keeling_2006_minim, dubois_2013_minim, gabelli_2013_shapin, Gaury_2014, glattli_2016_levit, baeuerle_2018_coher}. When the pulse carries noninteger multiples of a flux quantum, dynamical orthogonality catastrophe occurs, which can lead to electrons or holes with delocalized wave functions \cite{Yin2025b}. As the pulse flux increases by one flux quantum from zero, the dynamical orthogonality catastrophe is suppressed gradually, which can lead to a delocalization-to-localization transition for the individual electrons or holes. The transition can be described by a single-parameter scaling theory, where the scaling behavior can be fine-tuned by the long-time tail of the voltage pulse. The wave functions of the neutral eh pairs remains localized or delocalized when the flux is smaller than one flux quantum.

In this paper, we further show that the transition can also occur for the neutral eh pairs, when the pulse flux increases by more than one flux quantum. Unlike the case of individual electrons or holes, the wave functions of neutral eh pairs do not undergo a unidirectional transition as the flux increases. The transition can be a localization-to-delocalization transition or vice versa, which is controlled via the long-time tail of the voltage pulse. The localization-to-delocalization transition occurs in the case of short-tailed pulses, which decay faster than Lorentzian. The delocalization-to-localization transition occurs in the case of long-tailed pulses, which decay slower than Lorentzian. In both cases, the transitions for all the neutral eh pairs exhibit the same critical behavior, which can be seen by comparing their scaling functions and correlation lengths. No transition occurs in the case of Lorentzian pulses, as no localized neutral eh pair can be excited.

In contrast, the wave functions of individual electrons or holes always undergo a delocalization-to-localization transition as the pulse flux increases. The transitions for all electrons or holes exhibit the same critical behaviors in the case of short-tailed pulses, but exhibit different behaviors in the case of long-tailed and Lorentzian pulses. In the formal case, the directions of the transitions for the neutral eh pairs and individual electrons or holes are opposite. Certain localized neutral eh pairs can first evolve into delocalized ones, then split into individual electrons and holes with localized wave functions. This gives a reentrant localization, which has not been addressed so far. No reentrant localization occurs in the latter case. It is either because the directions of the transitions for the neutral eh pairs and individual electrons or holes are the same (long-tailed pulses), or because the transitions for neutral eh pairs is totally absent (Lorentzian pulses).

The paper is organized as follows. In Sec.~\ref{sec2}, we present our model for the charge injection and show how to extract the wave functions of electrons and holes from the scattering matrix. We introduce the inverse participation ratio (IPR) to characterize the localization/delocalization nature of the wave function. In Sec.~\ref{sec3}, we focus on transitions in the case of short-tailed pulses and show how does the reentrant localization occur. We perform the finite-size scaling analysis of the IPR to clarify the nature of the transitions. In Sec.~\ref{sec4}, we focus on the case of long-tailed and Lorentzian pulses, where the reentrant localization is absent. We summarized in Sec.~\ref{sec5}.

\section{Model and formalism\label{sec2}}

The model we used here is the same as that used in our previous works \cite{Yin2019, Yue_2019, Yin2025a, Yin2025b}. So we only briefly outline the main points
here. We consider the charge injection from a reservoir into a single-mode quantum conductor, which is driven by a time-dependent voltage $V(t)$ applied to the
electrode. We choose the driving voltage $V(t)$ of the form
\begin{eqnarray}
  V(t) = V_p(t_0+t) - V_p(t_0-t).
  \label{s2:eq10}
\end{eqnarray}
It corresponds to two successive pulses with the same shape but opposite signs, which are separated by a time interval $2t_0$. For simplicity, we assume the
pulse is symmetric, \ie, $V_p(t) = V_p(-t)$. We characterize the width of each pulse by the half width at half maximum $W$. The strength of the pulse can be
described by the flux $\varphi = (e/h) \int^{+\infty}_{-\infty} V_p(t) dt$. It is the Faraday flux of the voltage pulse normalized to the flux quantum $h/e$,
which is defined by the ratio of Planck’s constant $h$ and the elementary charge $e$. In the following part of the paper, we choose $e=h=W=1$.

The electron injection in this system can be fully characterized by the scattering matrix $\mathcal{S}(E, E')$ in the energy domain. It is only the function of
the energy difference $E-E'$ and hence can be written as $\mathcal{S}(E, E') = S(E-E')$, where
\begin{equation}
  S(E) = \int^{+\infty}_{-\infty} dt e^{2\pi i E t - i\phi(t)},
  \label{s2:eq20}
\end{equation}
with $\phi(t) = 2\pi \int^t_{-\infty} V(\tau) d\tau$ being the scattering phase due to the driving pulses. The many-body state of the injected electrons can be
expressed as
\begin{equation}
  | \Psi \rangle = \prod_{k=1, 2, 3, \dots} \Big[ \sqrt{1 - p_k} + i \sqrt{p_k} B^{\dagger}_e(k) B^{\dagger}_h(k) \Big] | F \rangle, \label{s2:eq30}
\end{equation}
where $|F\rangle$ represents the Fermi sea and $B^{\dagger}_e(k)$[$B^{\dagger}_h(k)$] represents the creation operator for the electron[hole] component of the
eh pairs. They can be expressed as
\begin{eqnarray}
  B^{\dagger}_e(k) & = & \int^{+\infty}_0 dE \psi^e_k(E) a^{\dagger}(E), \\
  B^{\dagger}_h(k) & = & \int^0_{-\infty} dE \psi^h_k(E) a(E).
                         \label{s2:eq40}
\end{eqnarray}
Note that as the emission is driven by a pair of pulse, electrons and holes are always emitted in pairs. 

The wave function $\psi^{e/h}_k(E)$ and excitation probability $p_k$ can be obtained from the polar decomposition of the scattering matrix. This can be done by
solving the following equation for $E>0$:
\begin{equation}
  \int^{+\infty}_{0} dE' S(E+E') \psi^{\ast}_{k}(E') = i \sigma \sqrt{p_k} \psi_k(E),
  \label{s2:eq50}
\end{equation}
with $\sigma = \pm 1$. The wave function of the electron $\psi^e_k(E)$ and hole $\psi^h_k(E)$ can be obtained as
$\psi^e_k(E) = \sigma \psi^h_k(-E) = \psi_k(E)$. In the numerical calculation, Eq.~(\ref{s2:eq50}) can be solved in the energy domain by using the singular
value decomposition \cite{Yin2019}.

We characterize the localization/delocalization nature of the wave function from the length-scale dependence of the inverse participation ratio (IPR). It can be
related to the time-domain wave function $\psi$ as
\begin{equation}
  P = \frac{1}{t_l} \int_{\rm{box\: origins}}  \sum_{\rm{box}(t_l)} \Big( \int_{t \in \rm{box}(t_l)} dt \left| \psi(t) \right|^2  \Big)^2, \label{s2:eq60}
\end{equation}
where we assume the wave function is normalized in the whole time domain $\int^{+t_{\rm max}/2}_{-t_{\rm max}/2} dt \left| \psi \right|^2 = 1$ with
$t_{\rm max} \to +\infty$. The symbol $\sum_{\rm{box}(t_l)}$ represents the summation over small boxes with a linear size $t_l$ into which one divides the whole
time domain $[-t_{\rm max}/2, +t_{\rm max}/2]$. The choice of the box origins is arbitrary, which gives different values of IPR. So the IPR is further averaged
over different box origins ($\frac{1}{t_l} \int_{\rm{box\: origins}} \dots$) to avoid this ambiguity.

One can distinguish the localized state from the delocalized one from the box size dependence of the IPR. When the box size $t_l$ is far from any characteristic
timescales, the IPR scales with $t_l$ as a power law: $P \sim t^{\tau}_l$. One has $\tau=0$ for the localized state and $\tau>0$ for the delocalized state,
where $\tau$ can be read from the slope of the IPR as a function of $t_l$ on a log-log scale. While the IPR for all the localized state approaches a constant,
the value of the constant depends on the temporal extent of the wave function. For example, if the wave function exhibits a single peak around $t=t_0$ so that
$|\psi(t)|^2=\delta(t-t_0)$, one has $P=1$ from Eq.~(\ref{s2:eq60}). If the wave function exhibits two peaks around $t=\pm t_0$ so that
$|\psi(t)|^2=(1/2)[\delta(t-t_0) + \delta(t+t_0)]$, one has $P=1/2$. This feature is helpful in characterizing the structure of the localized wave function.

The wave function and the excitation probability can be manipulated by changing the pulse profile. In particular, the tail of the pulse plays the dominant role
in the transition of the wave function \cite{Yin2025b}. To be more specific, we consider a family of pulses whose temporal profile can be written as
\begin{equation}
V_p(t) =  \frac{\varphi\Gamma(\alpha)}{\sqrt{\pi}\Gamma(\alpha-1/2)} \frac{1}{\left( t^2+1 \right)^{\alpha}},
\label{s2:eq70}
\end{equation}
with $\Gamma(x)$ representing the Gamma function. The parameter $\alpha$ satisfies $\alpha > 0.5$ so that the voltage pulse carries a finite flux
$\varphi$. The above expression corresponds to the Lorentzian pulse with an arbitrary exponent, which decays as $\sim t^{-2\alpha}$ at long
times. For wave functions of individual electrons or holes, we have shown that the transition exhibits different critical behaviors for short-tailed
($\alpha > 1.0$) and long-tailed ($\alpha < 1.0$) pulses. In this paper, we shall use these pulses to further explore the transition for the neutral eh
pairs and the related reentrant localization.

\section{Short-tailed pulses \label{sec3}}

\begin{figure}
  \includegraphics[width=1\linewidth]{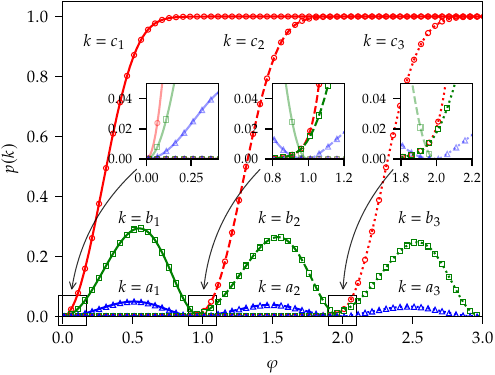}%fig_pk_PowA1.5_t0=32
  \caption{\label{fig-s3-10} The excitation probabilities $p(k)$ as a function of $\varphi$, corresponding to $\alpha=2.0$, $t_0=32$ and
    $\varphi \in [0, 3]$. There exists nine eh pairs, which are labeled by $a_n$, $b_n$ and $c_n$ with $n=1$, $2$ and $3$. The probabilities of the eh pairs $a_n$, $b_n$ and $c_n$ are shown by blue triangles, green squares and red circles, respectively. The probabilities corresponding to $n=1$, $2$ and $3$ are shown by solid, dashed and dotted curves. The three insets show the zoom-in around the point $\varphi=0$, $1$ and $2$, respectively.}
\end{figure}

We start our discussion from the case of short-tailed pulses, corresponding to $\alpha > 1.0$. First we find the dominated eh pairs from the excitation
probability. In Fig.~\ref{fig-s3-10}, we show the excitation probability as a function of the pulse flux $\varphi$, corresponding to $\alpha=2.0$, $t_0=32$, and
$\varphi \in [0, 3]$. In this case, the excitation is dominated by nine eh pairs. From the main panel, it seems that the probabilities are periodic. Each time
$\varphi$ is increased by one, three additional eh pairs emerge, which are labeled by $a_n$, $b_n$ and $c_n$ with $n=1$, $2$ and $3$. The probability of $c_n$
increases monotonically from $0.0$ to $1.0$ as $\varphi$ increases from $n-1$ to $n$. In contrast, the probabilities of $a_n$ and $b_n$ take their maximums when
$\varphi$ is close to the half-integer value $n-1/2$ and drop to zero when $\varphi$ approaches the integer values $n-1$ and $n$.

However, a close look around the point $\varphi=n-1$ shows that the probabilities of $b_n$ and $c_n$ for $n>1$ do not drop exactly to zero at this point. For
example, the probabilities of $b_2$ and $c_2$ remain nonzero and become degenerated when $\varphi$ drops below $1.0$. This is highlighted by the red and green dashed curves in the middle inset of Fig.~\ref{fig-s3-10}. A similar behavior can also be seen for the eh pair $b_3$ and $c_3$, as shown in the right inset corresponding to $\varphi=1$. In
contrast, no degeneracy occurs for the probabilities of $b_1$ and $c_1$, which all drop to zero for $\varphi=0$ [see the left inset]. Unlike $b_n$ and $c_n$, the probabilities of all $a_n$ behave in a similar way: All of them drop exactly to zero as $\varphi$ reaches the integer values $n-1$ and $n$.

Different behaviors of the probabilities suggest different natures of the wave functions. In our previous work \cite{Yin2025a, Yin2025b}, we have shown that the wave function of $a_1$ and $b_1$ are always delocalized, while the wave function of $c_1$ can undergo a delocalization-to-localization transition as $\varphi$ approaches $1.0$. From the behavior of the probabilities, one expect that the wave functions are still delocalized for the eh pair $a_n$ with $n>1$, but the wave functions of $b_n$ and $c_n$ can show different features. 

\subsection{Wave function and inverse participation ratio \label{sec3.1}}

\begin{figure}
\includegraphics[width=1\linewidth]{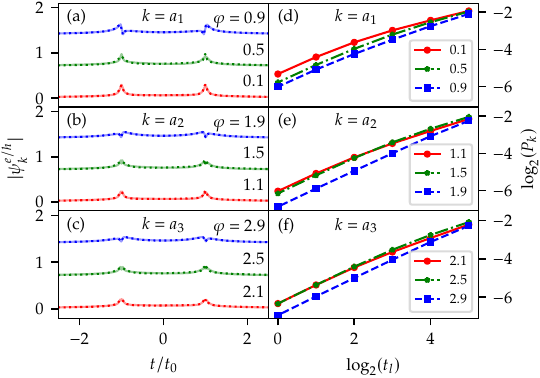}%fig_PowA1.5_Zq-Wf_t0=32_ax
\caption{\label{fig-s3-20} Wave functions of electrons (solid) and holes (dotted) for the eh pair $a_1$(a), $a_2$(b) and $a_3$(c) for three typical values of $\varphi$. Wave functions for different values of $\varphi$ are shifted vertically relative to each other for clarity. The corresponding IPR are shown alongside in (d-f). }
\end{figure}

To show this, we first compare the wave functions of the eh pairs $a_n$($n=1, 2, 3$) in Figs.~\ref{fig-s3-20}(a)-\ref{fig-s3-20}(c), where the solid and dotted curves represent the modulus of the wave functions for electrons and holes, respectively. One finds that the electron and hole wave functions are largely overlapped in the time domain. This indicates that $a_n$ are neutral eh pairs, which do not contribute to the net charge injection. The wave functions exhibit two peaks centralized around $t=\pm t_0$, which are rather broad. This indicates that the wave functions are delocalized. The delocalization nature can be characterized quantitatively by using the IPR, which are plotted alongside the wave functions in Figs.~\ref{fig-s3-20}(d)-\ref{fig-s3-20}(f). One can see that all the IPR increase almost linearly as a function of the box size $t_l$ on a log-log scale, which is a typical signature of the delocalized wave function. These results confirm that the eh pairs $a_n$ are all neutral eh pairs with delocalized wave functions.

\begin{figure}
  \includegraphics[width=1\linewidth]{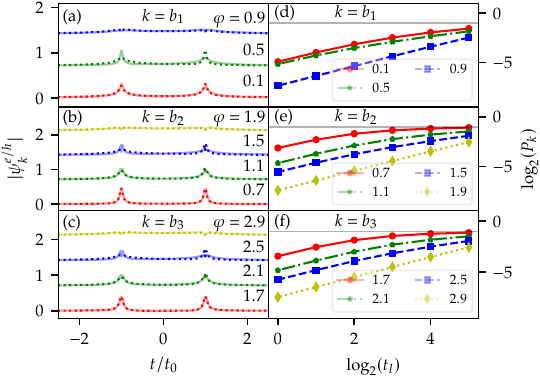}%fig_PowA1.5_Zq-Wf_t0=32_bx
  \caption{\label{fig-s3-30} The same as Fig.~\ref{fig-s3-20} but for the eh pair $b_1$, $b_2$ and $b_3$. The gray horizontal solid lines in (d-f) represents $P_k=1/2$. }
\end{figure}

Now we turn to the eh pairs $b_n$ and $c_n$, whose probabilities can be degenerated for $n>1$ when $\varphi$ drops below $n-1$. We have shown that the degeneracy indicates that
they are bonding and antibonding states built from neutral eh pairs localized around $t=\pm t_0$, respectively \cite{Yin2025a}. The wave functions of
$b_n$ does show the feature of the localized neutral eh pair when $\varphi$ is smaller than $n-1$: The electron and holes wave functions are largely
overlapped and both of them exhibits two-peak structures, which are well-localized around the point $t=\pm t_0$. This is illustrated in Figs.~\ref{fig-s3-30}(b)
and \ref{fig-s3-30}(c) by the red curves, corresponding to $b_2$ for $\varphi=0.7$ and $b_3$ for $\varphi=1.7$, respectively. The localization nature can be better seen from the IPR, which are shown by the red solid curves in Figs.~\ref{fig-s3-30}(e) and \ref{fig-s3-30}(f). The IPR of both wave functions tend to be independent on the box size $t_l$ when $t_l$ is large enough, which is a typical signature of the localized wave function. As the localized wave function exhibit two localized peaks, the IPR approaches the constant $1/2$, as we have explained in Sec.~\ref{sec2}. When $\varphi$ becomes larger than $n-1$, the two-peak structures in the wave functions remain unchanged, but
the peak width increases rapidly. This is illustrated in Figs.~\ref{fig-s3-30}(b) and \ref{fig-s3-30}(c) by the green, blue and yellow curves. The corresponding IPR increase almost linearly as a function of $t_l$ on a log-log scale when $t_l$ is large, as shown in Figs.~\ref{fig-s3-30}(e) and \ref{fig-s3-30}(f) by the green dash-dotted, blue dashed and yellow dotted curves. This corresponds to a neutral eh pair with delocalized wave functions. Hence $b_2$ and $b_3$ are neutral eh pairs, whose wave functions undergo a localization-to-delocalization transition as $\varphi$ increases from below to above the points $\varphi=1$ and $\varphi=2$, respectively. In contrast, no signature of the transition is found for the eh pair $b_1$, whose wave functions and IPR are shown in Fig.~\ref{fig-s3-30} (a) and (c). The wave functions and the IPR of $b_1$ show similar features as the ones of the eh pair $a_1$, indicating that it is always a neutral eh pair with delocalized wave functions. These results confirm that the eh pair $b_n$ show two different behaviors: 1) The wave function of $b_n$ for $n>1$ can undergo a delocalization-to-localization transition as $\varphi$ increases. 2) The wave function of $b_1$ is always delocalized.

\begin{figure}
  \includegraphics[width=1\linewidth]{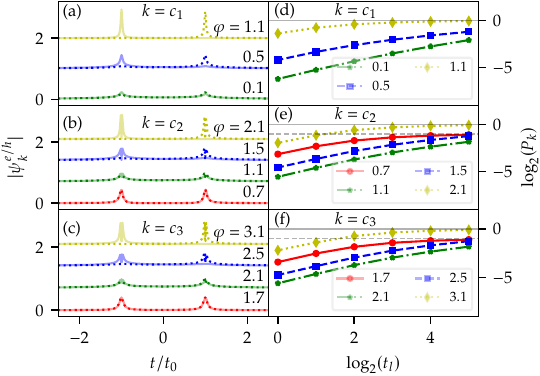}%fig_PowA1.5_Zq-Wf_t0=32_cx
  \caption{\label{fig-s3-40} The same as Fig.~\ref{fig-s3-20} but for the eh pair $c_1$, $c_2$ and $c_3$. The gray horizontal solid and dashed lines in (d-f) represent $P_k=1$ and $P_k=1$, respectively. }
\end{figure}

The difference can also be seen for the eh pair $c_n$, but it occurs in a different manner. First, one finds that all the eh pairs $c_n$ can undergo a
delocalization-to-localization transition as $\varphi$ increases from below to above $\varphi=n$. Take the eh pair $c_2$ for example: As $\varphi$ increases
from $1.1$ to $2.3$, the slope of the IPR decreases rapidly, as shown in Fig.~\ref{fig-s3-40}(e) by the green dash-dotted, blue dashed and yellow dotted
curves. This suggests the presence of a delocalization-to-localization transition as $\varphi$ increases. In the delocalization phase, the wave functions of the electron and hole are largely overlapped. Both of them exhibit two broad peaks, corresponding to a neutral eh pair with delocalized wave functions. This is demonstrated by the green curves in Fig.~\ref{fig-s3-40}(b), corresponding to $\varphi=1.1$. In the localization phase, the wave function of the electron and hole are well-separated, indicating that they are individually injected by the positive and negative pulses. They exhibit only a single localized peak, as illustrated by the yellow curves in Fig.~\ref{fig-s3-40}(b). The corresponding IPR approaches the constant $1$, as shown by the yellow dotted curve in Fig.~\ref{fig-s3-40}(e). These results suggest that, during the delocalization-to-localization transition, the eh pair $c_2$ evolves from a delocalized neutral eh pair into individual electrons and holes with localized wave functions. A similar transition can also be seen for the eh pair $c_1$ and $c_3$, as illustrated in Fig.~\ref{fig-s3-40}(a) and (d) for $c_1$ and Fig.~\ref{fig-s3-40}(c) and (f) for $c_3$. The transition has been studied in our previous works \cite{Yin2025a, Yin2025b}.

But the eh pair $c_n$ with $n>1$ can undergo another transition as $\varphi$ drops below $n-1$. For example, the wave function of $c_2$ exhibits two well-localized peaks for $\varphi=0.7$, while the corresponding IPR approaches the constant $1/2$. This is illustrated in Fig.~\ref{fig-s3-40}(b) and (e) by the red solid curve. This suggests the presence of another localization phase, where the wave function corresponds to a localized neutral eh pair. As $\varphi$ increases from below to above $1.0$, the localized neutral eh pair evolves into a delocalized one, which gives another localization-to-delocalization transition. The transition can also be seen for $c_3$ [Fig.~\ref{fig-s3-40}(c) and (f)], but is absent for $c_1$ [Fig.~\ref{fig-s3-40}(a) and (d)].

From the above discussion, one can see that, as the flux $\varphi$ increases, there exists two kinds of transitions: 1) A delocalization-to-localization transition, where a delocalized neutral eh pairs evolves into individual electrons and holes. This occurs for the all eh pair $c_n$ around the critical point $\varphi=n$. 2) A localization-to-delocalization transition, where a localized neutral eh pair evolves into a delocalized one. This occurs for the eh pair $c_n$ and $d_n$ around the critical point $\varphi=n-1$ for $n>1$. In particular, $c_n$ with $n>1$ can undergo both transitions as $\varphi$ increases, leading to the presence of a reentrant localization. While these transitions can be seen qualitatively from the behavior of the wave function and the IPR, a quantitative description requires a detailed finite-size scaling analysis. This shall be discussed in the following section.

\subsection{Finite-size scaling analysis \label{sec3.2}}

\begin{figure}
  \includegraphics[width=1\linewidth]{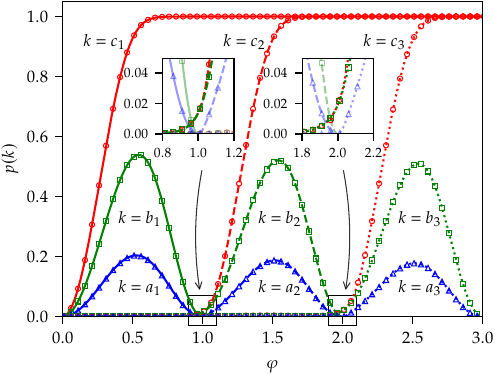}%fig_pk_PowA1.5_t0=512
  \caption{\label{fig-s3-45} The same as Fig.~\ref{fig-s3-10}, but for $t_0=512$. Only the zoom-in around $\varphi=1$ and $2$ are shown in the inset.}
\end{figure}

We choose IPR as the scaling variable for the finite-size scaling analysis. We assume the IPR for a given $t_0$ can be described by a single-parameter scaling
function $f(x)$ as
\begin{equation}
  P_c(t_l, \varphi) = f[t_l/\xi(\varphi)],
  \label{s3:eq10}
\end{equation}
with $\xi(\varphi)$ representing the correlation length. Both the scaling function and the correlation length can be obtained by using the data collapse method. To do this, we rescale the IPR for various $t_l$ and $\varphi$ so that all the data points can be collapsed into a single curve, corresponding to the scaling function $f(x)$. To perform the scaling analysis, the IPR should be evaluated for large $t_0$ to minimize finite-size effect. We choose $t_0=512$ and $t_{\rm max}=8192$, which is proved to be suitable in our previous works \cite{Yin2025a, Yin2025b}. For such a large value of $t_0$, the excitation probabilities of the delocalized eh pairs are enhanced, but the key features remains the same as the case for $t_0=32$. This can be seen by comparing Fig.~\ref{fig-s3-45} to Fig.~\ref{fig-s3-10}.

\begin{figure}
  \includegraphics[width=1\linewidth]{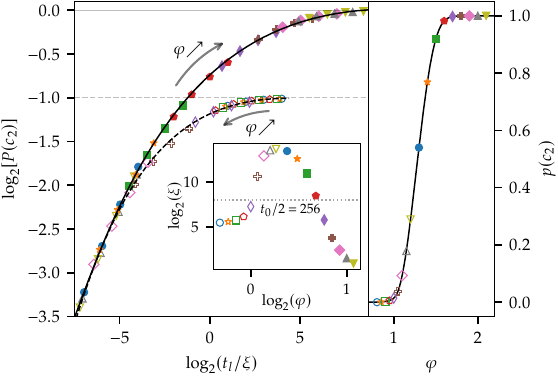}%fig_fss_PowA1.5_T0=512_c2
  \caption{\label{fig-s3-50} Left panel: The IPR $\log_2[P(c_2)]$ as a function of $\log_2(t_l/\xi)$, corresponding to $\alpha=2.0$, $t_0=512$ and $\varphi \in [0.8, 2.1]$. The black arrows shows the direction of increasing $\varphi$. The gray horizontal solid and dashed lines represent $P(c_2)=1$ and $P(c_2)=1/2$, respectively. The correlation length $\xi$ as a function of $\varphi$ is shown in the inset on a log-log scale. The gray horizontal dotted line represents $\xi=t_0/2=256$. Right panel: The excitation probability $p(c_2)$ as a function of $\varphi$. In both panels, data points with the same colors and markers correspond to the same values of $\varphi$. }
\end{figure}

Let us first focus on the eh pair $c_2$. The collapse of the IPR is demonstrated in the left panel of Fig.~\ref{fig-s3-50}. The correlation length are shown in the inset, while the corresponding excitation probabilities are shown in the right panel. Data points with the same colors and markers correspond to the same values of $\varphi$. By properly chosen $\xi$, one finds that the IPR can be collapsed into a curve with two different branches. The lower one (black dashed) corresponds to the IPR for $\varphi \in [0.8, 1.2]$, as shown by the unfilled markers. It represents the scaling function of the localization-to-delocalization transition around the critical point $\varphi=1.0$. As $\varphi$ drops below $1.0$, the IPR saturates to the constant $1/2$, indicating that the localization phase corresponds to a localized neutral eh pair with a two-peaked wave function. The upper one (black solid) corresponds to the IPR for $\varphi \in [1.3, 2.1]$, as shown by the filled markers. It represents the scaling function of the delocalization-to-localization transition around the critical point $\varphi=2$. As $\varphi$ goes above $2.0$, the IPR saturates to the constant $1$, indicating that the localization phase corresponds to individual electrons (holes) with single-peaked wave functions. The two-branch structure of the scaling function describes the reentrant localization as $\varphi$ increases from below $1.0$ to above $2.0$. The reentrant localization can also be seen from the correlation length, which is shown on a log-log scale in the inset of the left panel. The correlation length exhibits a non-monotonic behavior as $\varphi$ increases. It remains finite when $\varphi$ is smaller than $1.0$ or larger than $2.0$, corresponding to the two localization phases. In contrast, it diverges when $\varphi$ is far from the two critical points, corresponding to the delocalization phase. 

\begin{figure}
  \includegraphics[width=1\linewidth]{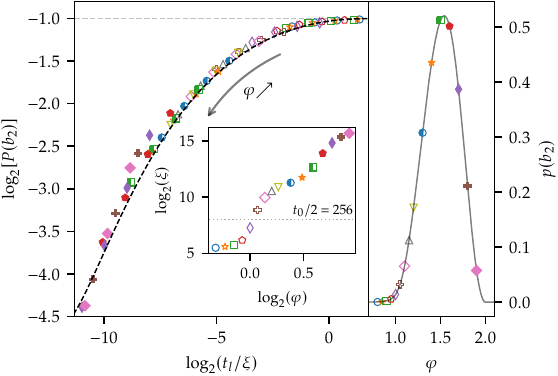}%fig_fss_PowA1.5_T0=512_b2
  \caption{\label{fig-s3-60} The same as Fig.~\ref{fig-s3-50} but for the eh pair $b_2$. }
\end{figure}

Now we turn to the eh pair $b_2$. The collapse of the rescaled IPR is demonstrated in the left panel of Fig.~\ref{fig-s3-60}. Only a single localization phase can be identified, corresponding to $\varphi < 1.0$. The IPR saturates to the constant $1/2$ in the localization phase, corresponding to a localized neutral eh pair with two-peaked wave functions. As $\varphi$ goes above $1.0$, the localization-to-delocalization transition occurs, whose scaling function is illustrated by the black dashed curve. The scaling function is obtained by fitting the IPR around the critical point ($\varphi \in [0.8, 1.2]$), which are shown in the left panel by the unfilled markers. It can give a good estimation of the IPR when $\varphi$ is not very far from the critical point, as illustrated by the half-filled markers, corresponding to $\varphi \in (1.2, 1.5)$. As $\varphi$ goes above $1.5$, the IPR depart from the fitted scaling function, as illustrated by the filled markers. The correlation length also diverges differently when $\varphi$ is far from the critical point. This can be seen by comparing the filled markers to the half-filled and unfilled ones in the inset. So although the reentrant localization is absent for the eh pair $b_2$, the IPR can still show a different scaling behavior when $\varphi$ approaches $2.0$.

\begin{figure}
  \includegraphics[width=1\linewidth]{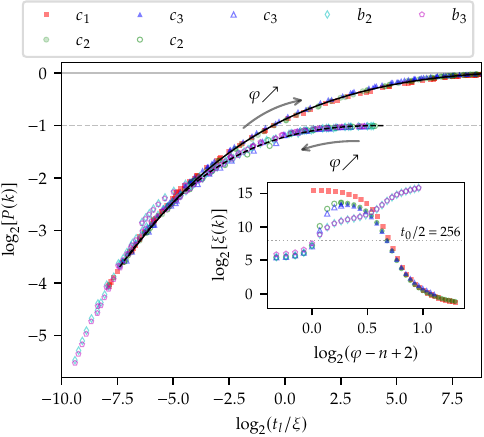}%fig_fss_PowA1.5_T0=512_cd-unif
  \caption{\label{fig-s3-70} The IPR $\log_2[P(k)]$ as a function of $\log_2(t_l/\xi)$ for different eh pair $k=b_n$ and $k=c_n$ with $n=1$, $2$ and $3$. For $c_2$ and $c_3$, the IPR corresponding to localization-to-delocalization (delocalization-to-localization) transition are demonstrated by the unfilled (filled) markers. The black arrows show the direction of increasing $\varphi$. The gray horizontal solid and dashed lines represent $P(k)=1$ and $P(k)=1/2$, respectively. The correlation length $\xi$ as a function of $\varphi$ is shown in the inset on a log-log scale. The gray horizontal dotted line represents $\xi=t_0/2=256$.}
\end{figure}

The transitions for other eh pairs can be obtained in a similar way. One may wonder if delocalization-to-localization (or localization-to-delocalization) transitions for different eh pairs exhibit the same critical behavior. To see this, we compare their scaling functions and correlation lengths in Fig.~\ref{fig-s3-70}. The IPR and scaling function are demonstrated in the main panel. The filled markers represent the IPR corresponding to the delocalization-to-localization transitions for the eh pair $c_n$ around the critical point $\varphi=n$ with $n=1, 2, 3$. The unfilled markers represent the IPR corresponding to the localization-to-delocalization transitions for the eh pair $c_n$ and $d_n$ around the critical point $\varphi=n-1$ with $n=2, 3$. In both cases, the IPR for different eh pairs can be collapsed into the same curve (black solid or dashed), indicating that the corresponding transition can be described by the same scaling function. The correlation lengths $\xi$ are plotted on a log-log scale in the inset. Note that we plot $\xi$ as a function of $\varphi-n+2$ so that the critical points for different eh pairs $c_n$ and $d_n$ coincide in the figure. In doing so, one finds that all the correlation lengths also diverge in a similar manner for $\xi < t_0/2$. The divergence show different behaviors for $\xi > t_0/2$, which can be attributed to the finite-size effect. These results show that all the delocalization-to-localization (localization-to-delocalization) transitions can be described by the same scaling function and correlation length, and hence they exhibit the same critical behavior.

\section{long-tailed pulses \label{sec4}}

\begin{figure}
  \includegraphics[width=1\linewidth]{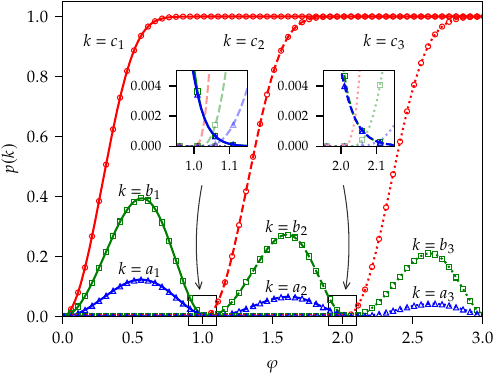}%fig_pk_PowA035_t0=512
  \caption{\label{fig-s4-10} The same as Fig.~\ref{fig-s3-10}, but for $\alpha=0.85$ and $t_0=512$. Only the zoom-in around $\varphi=1$ and $2$ are shown in the inset. }
\end{figure}

Now we turn to the case of long-tailed pulses, corresponding to $\alpha < 1.0$. In Fig.~\ref{fig-s4-10}, we show the excitation probability as a function of the
pulse flux $\varphi$ for $\alpha=0.85$ and $t_0=512$. By comparing the main panels of Fig.~\ref{fig-s3-45} and Fig.~\ref{fig-s4-10}, it seems that probabilities
exhibit qualitatively similar behaviors for both the short-tailed ($\alpha=2.0$) and long-tailed ($\alpha=0.85$) pulses. But there exists crucial differences when the
flux $\varphi$ is close to an integer. This can be better seen from the zoom-in around the critical point $\varphi=n$. Take, for example, the critical point
$\varphi=1$. The zoom-in of the probabilities around this point are demonstrated in the left inset of Fig.~\ref{fig-s4-10}. First, one finds that the
probability of $b_2$ and $c_2$ drops to zero for $\varphi<1$, as illustrated by the red and green dashed curve in the left inset of Fig.~\ref{fig-s4-10}. This is distinctly different from the case of short-tailed pulses, where the probability of $b_2$ and $c_2$ remains nonzero for $\varphi<1$ [see the red and green dashed curves in the left inset of Fig.~\ref{fig-s3-45}]. Second, the probabilities of $a_1$ and $b_1$ remains nonzero and become degenerated as $\varphi$ increases from below to above $1.0$. This is highlighted by the green and blue solid curves in the left inset of Fig.~\ref{fig-s4-10}. This is also different from the case of short-tailed pulses, where the probabilities of $a_1$ and $b_1$ drop to zero for $\varphi > 1.0$. Similar differences can also be seen from the right inset, corresponding to the point $\varphi=2.0$. These behaviors suggest that the transition of the all eh pairs $a_n$, $b_n$ and $c_n$ can exhibit different behaviors in the case of long-tailed pulses.

\begin{figure}
  \includegraphics[width=1\linewidth]{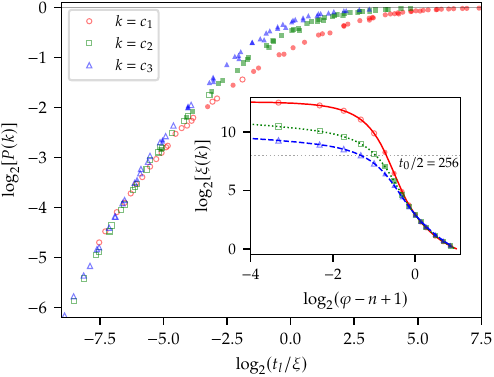}%fig_fss_PowA035_t0=512_cx
  \caption{\label{fig-s4-20} The IPR $\log_2[P(k)]$ as a function of $\log_2(t_l/\xi)$ for different eh pair $k=c_n$, with $n=1$ (red dots), $n=2$ (green squares) and $n=3$ (blue triangles), corresponding to $\alpha=0.85$ and $t_0=512$. Data points with filled(unfilled) markers correspond to the flux $\varphi$ near(far from) the critical point $\varphi=n$. The gray horizontal line in the main panel corresponds to $P(k)=1$. The gray horizontal dotted line represents $\xi=t_0/2=256$.}
\end{figure}

As the probabilities of $b_n$ and $c_n$ ($n>1$) drops to zero for $\varphi<n-1$, the localization phase below this point vanishes. As far as $c_n$ is concerned, only the localization phase above the point $\varphi=n$ is preserved and hence only the delocalization-to-localization transition occurs. The corresponding scaling function has only a single branch. This can be seen from the collapse of the IPR shown in the main panel of Fig.~\ref{fig-s4-20}. In this case, the IPR for different $c_n$ are collapsed into different curves, corresponding to different scaling functions. The correlation lengths also diverge in different ways, as shown in the inset. This indicates the critical behavior is different for different $c_n$.

\begin{figure}
  \includegraphics[width=1\linewidth]{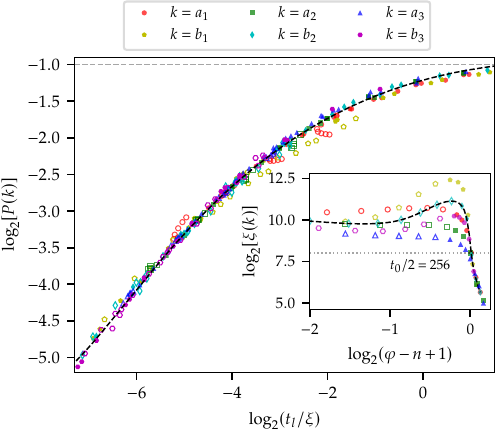}%fig_fss_PowA035_t0=512_abx
  \caption{\label{fig-s4-30} The IPR $\log_2[P(k)]$ as a function of $\log_2(t_l/\xi)$ for different eh pair $k=a_n$ and $b_n$ with $n=1$, $2$ and $3$. The correlation lengths are shown in the inset. Data points with filled(unfilled) markers correspond to the flux $\varphi$ near(far from) the critical point $\varphi=n$. }
\end{figure}

While the localization phase below $\varphi<n-1$ also vanishes for $b_n$, another localization phase emerges for $\varphi>n$. The localization phase also emerges for $a_n$. This makes both the eh pair $a_n$ and $b_n$ undergo delocalization-to-localization transitions around the critical point $\varphi=n$. The collapse of the IPR are shown in the main panel of Fig.~\ref{fig-s4-30}. In the vicinity of the critical point, we find that all the IPR (filled markers) can be collapsed approximately into the same black dashed curve, corresponding to the same scaling function. The correlation length also diverges in a similar way when $\xi$ is smaller than $t_0/2$, which can be seen from the inset. This suggests the transition exhibit similar critical behaviors. The IPR and correlation length show a large departure from each other when $\varphi$ goes far from the critical point, as illustrated by the unfilled markers. We expect this is due to the finite-size effect as $\xi$ becomes much larger than $t_0/2$ in this region.

\begin{figure}
  \includegraphics[width=1\linewidth]{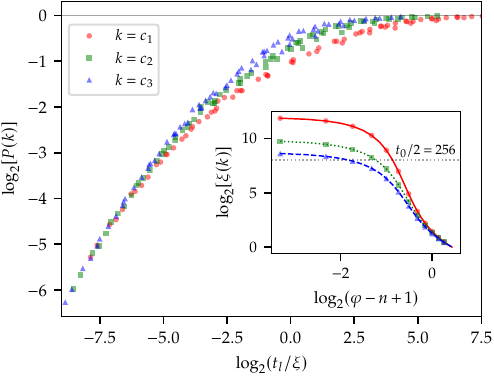}%fig_fss_Lor_t0=512_cx
  \caption{\label{fig-s4-40} The same as Fig.~\ref{fig-s4-20}, but for the Lorentzian pulse ($\alpha=0.5$).}
\end{figure}

What happens in the case of Lorentzian pulses? As the excitation of localized eh pairs are totally absent, there exists no localization transition for the neutral eh pairs $a_n$ and $b_n$. The reentrant localization is also absent for $c_n$. The only transition left is the delocalization-to-localization transition for $c_n$ around the critical point $\varphi=n$. The collapse of the IPR and the correlation lengths are shown in Fig.~\ref{fig-s4-40}. One finds that they exhibit different behaviors for different $c_n$, which is similar to the case of long-tailed pulses [Fig.~\ref{fig-s4-20}].

As the transition exhibit quite different behaviors in the case of long-tailed and short-tailed pulses, one may wonder what happens when the pulse evolves from a long-tailed one into a short-tailed one. To show this, we focus on two representative eh pairs $c_2$ and $b_2$. The formal one can evolve from a neutral eh pair into individual electron and hole, while the latter one is always a neutral eh pair.

\begin{figure}
  \includegraphics[width=1\linewidth]{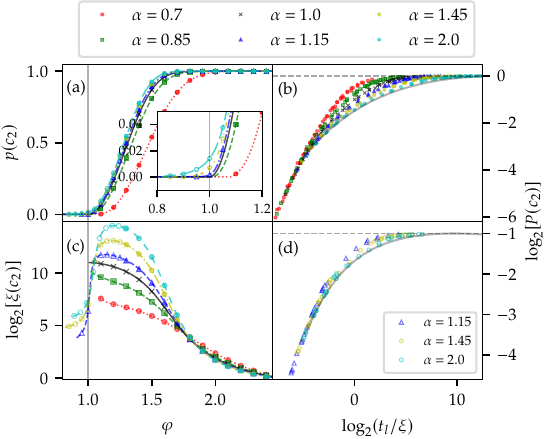}%fig_fss_PowAmn_t0=512_c2
  \caption{\label{fig-s5-10} (a) The excitation probability $p(c_2)$ as a function of flux $\varphi$ for six typical values of $\alpha$. The inset show the zoom-in around the point $\varphi=1.0$.  (b) The IPR $\log_2[P(c_2)]$ as a function of $\log_2(t_l/\xi)$ for the delocalization-to-localization transition around the point $\varphi=2.0$. The gray horizontal dashed curve represents $P(c_2)=1.0$. (c) The correlation length as a function of $\varphi$ on a semi-log scale. (d) The IPR $\log_2[P(c_2)]$ as a function of $\log_2(t_l/\xi)$ for the localization-to-delocalization transition around the point $\varphi=1.0$. The gray horizontal dashed curve represents $P(c_2)=1/2$. The black vertically lines in (a) and (c) corresponds to $\varphi=1.0$. The black solid curves in (b) and (d) represents the short-tailed pulse limit, which is obtained from the IPR with $\alpha=4.0$. In all the figures, data points with the same colors and markers correspond to the same values of $\varphi$. }
\end{figure}

Let us first focus on the eh pairs $c_2$. The excitation probability and correlation lengths are demonstrated in Fig.~\ref{fig-s5-10}(a) and (c), corresponding to six typical values of $\alpha$ from $0.7$ up to $2.0$. As $\alpha$ goes from below to above $1.0$ (Lorentzian), the zero of the probability $p(c_2)$ moves from the region $\varphi>1.0$ into the region $\varphi<1.0$, as can be seen from the inset of Fig.~\ref{fig-s5-10}(a). In the meantime, the localization phase in $\varphi<1.0$ emerges, where the correlation lengths drop to finite values for $\alpha=1.15$, $\alpha=1.45$ and $\alpha=2.0$. This is shown in Fig.~\ref{fig-s5-10}(c), where we plot the correlation length as a function of $\varphi$ on a semi-log scale.

The collapse of the IPR for the delocalization-to-localization transition around $\varphi=2.0$ is demonstrated in Fig.~\ref{fig-s5-10}(b). While the IPR for different $\alpha$ are collapsed into different curves, they approach the same limit as $\alpha$ increases well above $1.0$. The limited case is demonstrated by the black solid curve, which is fitted from the IPR with $\alpha=4.0$ \footnote{For the clarification of the figure, their IPR are not shown.}. The collapse of the IPR for the localization-to-delocalization transition around $\varphi=1.0$ is demonstrated in Fig.~\ref{fig-s5-10}(d), which only exists in the case of short-tailed pulses. The IPR are also collapsed into different curves for different $\alpha$, but it is less sensitive to the value of $\alpha$. As $\alpha$ increases, they approach the short-pulse limit (black solid curve), which is also obtained from the IPR with $\alpha=4.0$.

\begin{figure}
  \includegraphics[width=1\linewidth]{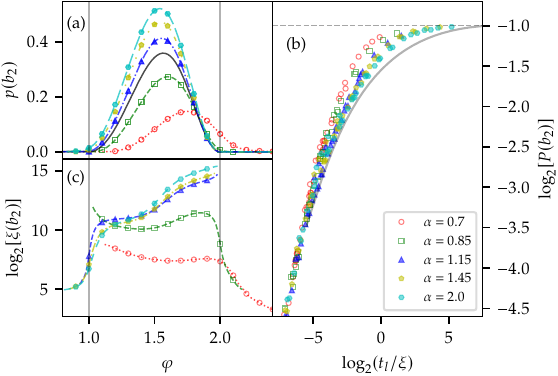}%fig_fss_PowAmn_t0=512_b2
  \caption{\label{fig-s5-20} (a) The excitation probability $p(b_2)$ as a function of flux $\varphi$ for six typical values of $\alpha$. (b) The IPR $\log_2[P(b_2)]$ as a function of $\log_2(t_l/\xi)$ for the localization-to-delocalization(delocalization-to-localization) transition around the point $\varphi=1.0$($\varphi=2.0$). The gray horizontal dashed curve represents $P(b_2)=1/2$. (c) The correlation length as a function of $\varphi$ on a semi-log scale. The black vertically solid and dashed lines in (a) and (c) corresponds to $\varphi=1.0$ and $\varphi=2.0$, respectively. The black solid curves in (b) and (d) represents the short-tailed pulse limit, which is obtained from the IPR with $\alpha=4.0$. In all the figures, data points with the same colors and markers correspond to the same values of $\varphi$. The unfilled markers corresponds to the long-tailed pulses, while the filled markers corresponds to the short-tailed pulses. }
\end{figure}

Now we turn to the neutral eh pair $b_2$. There exist two zeros in the excitation probability of $b_2$, as illustrated in Fig.~\ref{fig-s5-20}(a). In the case of long-tailed pulses, the left zero is larger than $1.0$, while the right zero is larger than $2.0$, as illustrated by the red circles ($\alpha=0.7$) and green squares ($\alpha=0.85$). In these cases, only the localization phase for $\varphi>2.0$ exists, where the correlation lengths drops to finite values, as illustrated by the red circles ($\alpha=0.7$) and green squares ($\alpha=0.85$) in Fig.~\ref{fig-s5-20}(c). As $\alpha$ increases above $1.0$ (Lorentzian), the left zero drops below $1.0$, while the right zero drops below $2.0$, corresponding to the case of short-tailed pulses. Now the localization phase for $\varphi>2.0$ vanishes, while the localization phase for $\varphi<1.0$ emerges. Accordingly, the correlation length drops to finite values for $\varphi<1.0$, as shown by the blue triangles ($\alpha=1.15$), yellow pentagons ($\alpha=1.45$) and cyan hexagons ($\alpha=2.0$) in Fig.~\ref{fig-s5-20}(c). The collapse of the IPR is demonstrated in Fig.~\ref{fig-s5-20}(b). While the IPR with different $\alpha$ are collapsed into different curves, they approach the same short-tailed limit (black solid curve) as $\alpha$ increases. 

\section{Conclusion \label{sec5}}

In this paper, we investigate the localization transition of neutral eh pairs injected by a voltage pulse with arbitrary flux quantum. We find that their wave functions do not undergo a unidirectional transition as the flux increases. The transition can be a localization-to-delocalization transition or vice versa, which is controlled via the long-time tail of the voltage pulse. The localization-to-delocalization transition occurs in the case of short-tailed pulses, which decay faster than Lorentzian. The delocalization-to-localization transition occurs in the case of long-tailed pulses, which decay slower than Lorentzian. No transition occurs in the case of Lorentzian pulse, as
no localized neutral eh pair can be excited. In contrast, the wave functions of individual electrons or holes always undergo a delocalization-to-localization transition as the flux increases. In the case of short-tailed pulses, the directions of the transitions for the neutral eh pairs and individual electrons or holes are
opposite. Certain localized neutral eh pairs can first evolve into delocalized ones, then split into individual electrons and holes with localized wave functions, which gives a reentrant localization. The reentrant localization does not occur in the case of long-tailed pulses, where the directions of the two transitions are the
same. It is also absent in the case of Lorentzian pulse where the localized neutral eh pairs cannot be excited at all.

\begin{acknowledgments}
This work was partially supported by the National Key Research and Development Program of China under Grant No. 2022YFF0608302 and SCU Innovation Fund under Grant No. 2020SCUNL209.
\end{acknowledgments}

\bibliography{refs}{}
\end{document}